\documentclass[
 reprint,showkeys,showpacs,preprintnumbers,
 amsmath,amssymb,
 aps,
 prl,
 longbibliography,
 lengthcheck,%
]{revtex4-1}

\usepackage{graphicx}
\usepackage{bm}
\usepackage{amsmath}
\usepackage{dcolumn}
\usepackage{slashbox}

\newcommand{\reffig}[1]{Fig.\ref{#1}}
\newcommand{\refeq}[1]{Eq.(\ref{#1})}

\begin{document}

\title{Evaluating Laguerre-Gaussian beams with an invariant parameter}

\author{Junling Long, Ruifeng Liu, Yunlong Wang, Feiran Wang,  Pei Zhang,$^{\ast}$ Hong Gao, and Fuli Li}

\address{
MOE Key Laboratory for Nonequilibrium Synthesis and
Modulation of Condensed Matter, \\ Department of Applied Physics, Xi'an
Jiaotong University, Xi'an 710049, People's Republic of China
\\
$^*$Corresponding author: zhangpei@mail.ustc.edu.cn
}

\begin{abstract}We define a new parameter about Laguerre-Gaussian (LG) beams, named $Q^{l}_{p}$, which is only related to mode indices $p$ and $l$. This parameter is able to both evaluate and distinguish LG beams. The $Q^{l}_{p}$ values are first calculated theoretically and then measured experimentally for several different LG beams.
Another mode quality parameter, $ M^{2} $ value, is also measured. The comparison between $Q^{l}_{p}$ and $ M^{2}$ shows same trend for the quality of LG mode, while the measurement of $Q^{l}_{p}$ is much easier than $ M^{2}$.
\end{abstract}
\maketitle

Vortex beam attracts many interests for carrying orbital angular momentum (OAM),\textit{ i.e.}, the vortex beam with an azimuthal phase structure $exp(il\phi)$ carries OAM of $l\hbar$ per photon \cite{Al}. Light beams possessing OAM of $l$$\hbar$, where $l$ can take any integer value, have a potential ability to carry a large information since the unlimited range of $l$ forms an unbounded Hilbert space \cite{GM, GG}. One typical example of vortex beams is high-order Laguerre-Gaussian (LG) mode, which is one of transverse mode solutions for laser cavities. 
Practical generation of LG modes is always achieved using cylindrical lens mode converter \cite{MDC}, spiral phase plate \cite{PP} or fork hologram \cite{fork1,fork2}. Many applications of beams with OAM have been already demonstrated, from optical tweezers \cite{OT} to higher-dimensional quantum information encoding \cite{QIE}.

To obtain accurate result of an experiment with LG modes, it would be better to evaluate LG beams first. By now, a standard parameter, $M^{2}$ parameter \cite{M2}, can make it. The $M^{2}$ value is the ratio of the beam parameter product of an actual beam to that of an ideal Gaussian beam at the same wavelength. It can be used to quantify the degree of variation how the actual beam is different from such an ideal beam. The $M^{2}$ value can also sort the OAM of LG mode according to $ M^{2}  = 2p+l+1$ ($p$ and $l$ are the mode indices). However, credible measurement of the $M^{2}$ value need to record many intensity cross-sections along the propagation direction, especially around the waist and far field \cite{ISO}. So, despite the measurement hardness, $M^{2}$ parameter will also be invalid to evaluate the beam's quality or sort its OAM if only a cross-section intensity picture of an LG beam is given. Recently, a new way of making Fourier transform of the cross-section intensity picture can sort the beam's OAM  \cite{intensity}. But this method is not very accurate and cannot give an evaluation of the quality of LG modes. In this letter, we define a new parameter for LG beams which can both evaluate the quality of the LG mode and sort the OAM through only one cross-section intensity picture. The parameter, called $Q^{l}_{p}$ value ($p$ and $l$ are mode indices of LG beams), is firstly proved to be independent to the size or the observation position of the LG beam and only correlated with mode indices $p$ and $l$. Then, we carry out an experiment to demonstrate the theory. To make a comparison, the $M^{2}$ parameter is also measured.

The intensity of LG beams in the x-axis can be given by:
\begin{align}
{I} &={u^{l}_{p}}^{*}{u^{l}_{p}}\nonumber\\
&=\frac{2p!}{\pi(p+|l|)!}\frac{1}{\omega^{2}}(2(x/\omega)^{2})^{|l|}[L^{|l|}_{p}(2(x/\omega)^{2})]^{2}e^{-2(x/\omega)^{2}}\nonumber\\
& = \frac{2p!}{\pi(p+|l|)!}\frac{1}{\omega^{2}}{f^{l}_{p}(x/\omega)},
\label{ix}
\end{align}
where $u^{l}_{p}$ is the electromagnetic field amplitude, $L^{l}_{p}$ is a generalised Laguerre polynomial, $p$ is the radial mode index, $l$ is the topological charge, $k=2\pi/\lambda$, $\omega=\omega_{0}\sqrt{1+(z/z_{R})^{2}}$, $R=z(1+(z_{R}/z)^{2})$, $z_{R}=\frac{2\pi}{\lambda}$, and $f^{l}_{p}(x)=(2x^{2})^{|l|}[L^{|l|}_{p}(2x^{2})]^{2}e^{-2x^{2}}$.

\begin{figure}[htb1]
\centerline{\includegraphics[width=7.5cm]{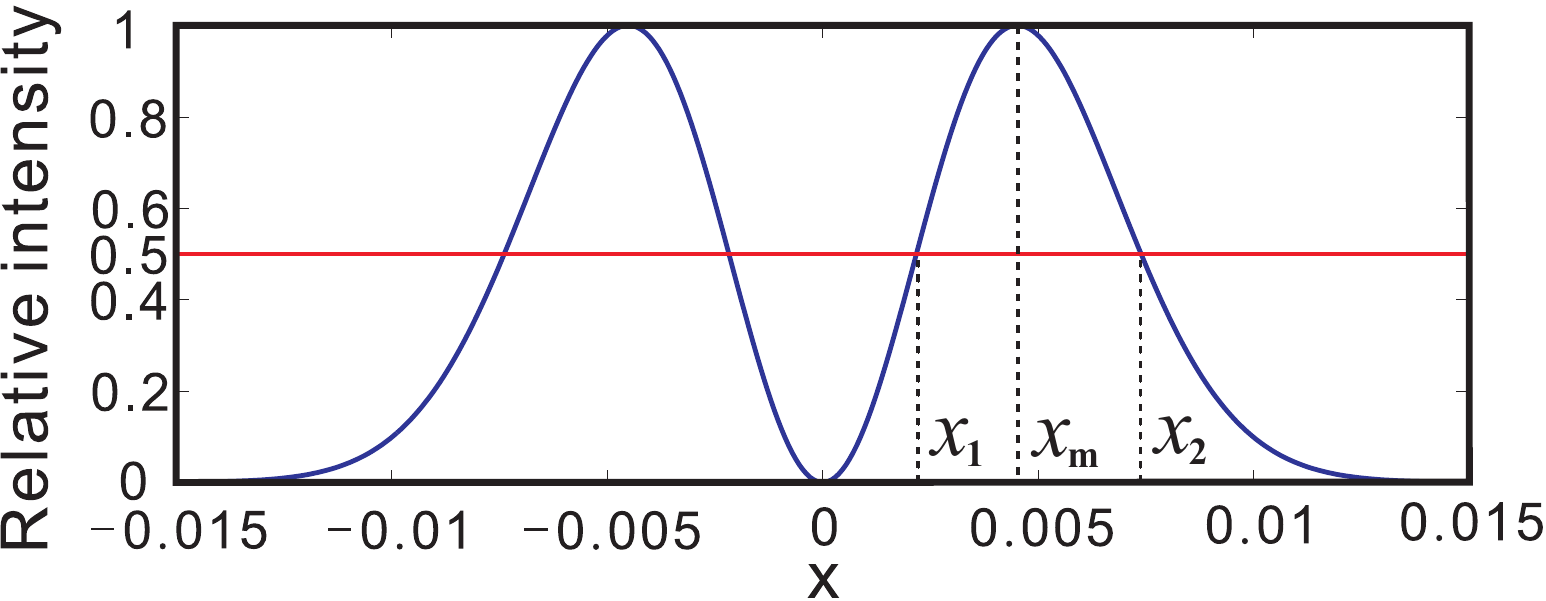}}
\caption{The relative intensity distribution of LG beams with $p=0$, $l=1$.}
\label{theory}
\end{figure}

\reffig{theory} shows the relative cross-section intensity of a beam with $l=1$ and $p=0$. $I_{m}$ is the first maximum value of function $I$ in the positive part of the x-axis and $I_{half}$ is equal to half of the $I_{m}$. In \reffig{theory}, $x_{1}$ and $x_{2}$ are the first two points on the x-axis where the intensity function $I$ reaches $I_{half}$, and $x_{m}$ is the point of $I$ reaching $I_{m}$.

We define $Q^{l}_{p}$ from the intensity function \emph{I}:
\begin{align}
{Q^{l}_{p}} &=\frac{x_{1}}{x_{2}-x_{1}}.
\end{align}
Let $t_i$ equals $x_{i}/\omega$, $i\in \{1,2,m\}$. So $Q^{l}_{p}$ can be calculated by  $Q^{l}_{p} = t_{1}/(t_{2}-t_{1})$. Next, we need to prove that $t_{1}$ and $t_{2}$ are only associated with mode indices $p$ and $l$.

The maximum value $I_m$ can be obtained from
\begin{align}
\frac{d{\emph{I}}}{d{\emph{x}}} &=\frac{2p!}{\pi(p+|l|)!}\frac{1}{\omega^{2}}\frac{d{f^{l}_{p}(x/\omega)}}{d(x/\omega)}\frac{1}{\omega} = 0.
\label{im}
\end{align}
And from \refeq{im}, it is obvious that $t_{m}$ can be calculated from:
 \begin{align}
 \frac{d{f^{l}_{p}(x/\omega)}}{d(x/\omega)} = \frac{d{f^{l}_{p}(t)}}{dt} = 0.
 \label{tm}
 \end{align}
Because the function $f^{l}_{p}(t)$ just depends on mode indices $p$ and $l$, $t_{m}$ obtained from \refeq{tm} is also only related to $p$ and $l$.

Then, $t_{1}$ and $t_{2}$ can be calculated from the following equation:
\begin{align}
\frac{1}{2} &=  {\frac{\emph{$I_{half}$}}{\emph{$I_{m}$}}} =\frac{\frac{2p!}{\pi(p+|l|)!}\frac{1}{\omega^{2}}f^{l}_{p}(x/\omega)}{\frac{2p!}{\pi(p+|l|)!}\frac{1}{\omega^{2}}f^{l}_{p}(x_{m}/\omega)} = \frac{f^{l}_{p}(t)}{f^{l}_{p}(t_{m})}.
\label{final}
\end{align}
Considering that $t_{m}$ and the form of the function $f^{l}_{p}(t)$ are both just associated with mode indices $p$ and $l$, we can conclude that $t_{1}$ and $t_{2}$ calculated from \refeq{final} is just related to $p$ and $l$. Hence, when mode indices $p$ and $l$ of LG beams are given, the $Q^{l}_{p}$ value calculated by  $Q^{l}_{p} = t_{1}/(t_{2}-t_{1})$ is unique.

\begin{figure}[htb]
\centerline{\includegraphics[width=7.5cm]{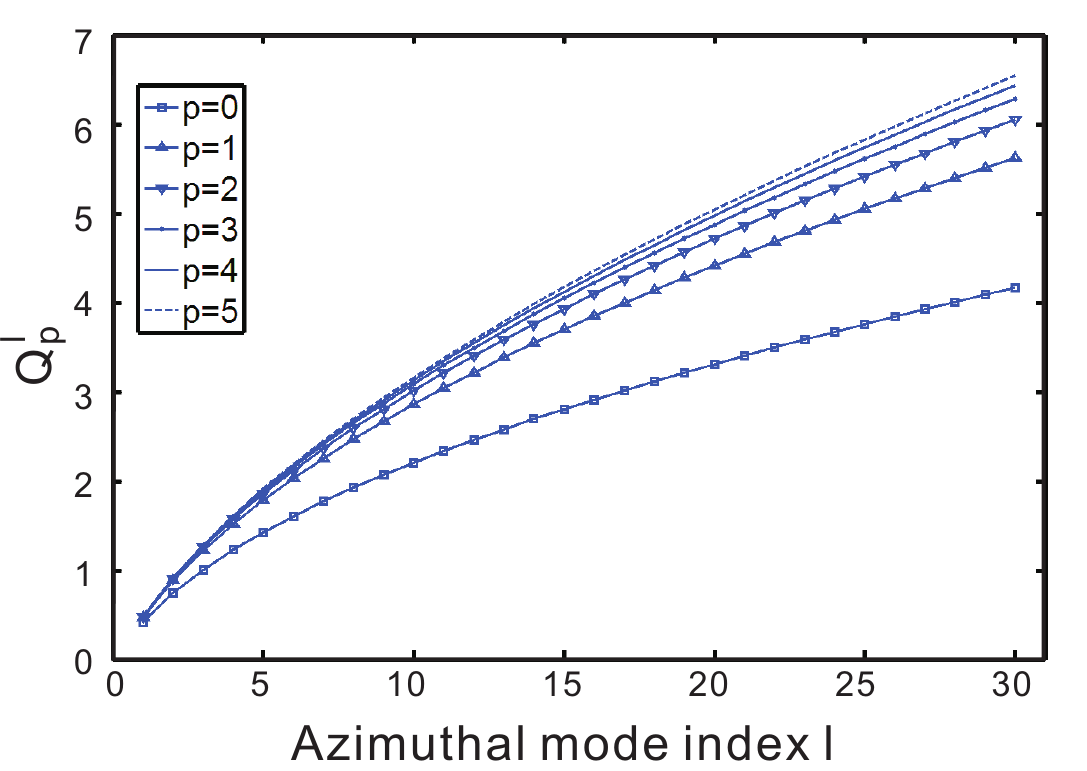}}
\caption{Theoretical $Q^{l}_{p}$ values of LG beams with different $p$ and $l$. }
\label{NC}
\end{figure}

It is obvious that \refeq{final} has no analytical solution. However we can we can numerically solve it. Some of the $Q^{l}_{p}$ values calculated numerically by computer are listed in Table I.
\begin{table}[h]
  \centering
  \caption{Some numerically calculated $Q^{l}_{p}$ values}\begin{tabular}{ccccc} \\ \hline
    \backslashbox{$p$} {$l$} & 1 & 2 & 3 & 4\\ \hline
    0 & 0.4170 & 0.7482 & 1.0093 & 1.2316  \\
    1 & 0.4754 & 0.8877 & 1.2265 & 1.5219  \\
    2 & 0.4821 & 0.9081 & 1.2632 & 1.5757 \\ \hline
  \end{tabular}
\end{table}
More theoretical $Q^{l}_{p}$ values are presented in \reffig{NC}, from which we can know that the $Q^{l}_{p}$ value is a strictly increasing function corresponding to $l$ when $p$ is fixed. So, $Q^{l}_{p}$ value of LG beam can be used to evaluate the mode quality and estimate OAM number.
Experimentally, we can take just one picture at any distance because the $Q^{l}_{p}$ is independent with $z$.
To measure the OAM of a high-order LG beam, the mode index $p$ of a LG beam can be firstly determined by the radial structure because the intensity cross-section consists of $p+1$ concentric rings. After the mode index $p$ of a LG beam is known, we can calculate $Q^{l}_{p}$ of the beam and then the mode index $l$ can be estimated by the one-to-one relationship between the $Q^{l}_{p}$ value and $l$.

To demonstrate the validity of this method, we experimentally generated some high-order LG beams and measured their $Q^{l}_{p}$ values.
In this experiment, computer-generated holograms were used to produce high-order LG modes with $l$ ranging from 1 to 3 and $p=0$ (pictures are shown in the middle of \reffig{result}). A charge coupled device (CCD) was used to record the cross-section intensity. Particularly worth mentioning is that the CCD camera must respond linearly to the light intensity. A power meter was used to test and verify the linear response of CCD camera to the light intensity in our experiment.

For every mode, we recorded a far-field intensity pattern every 10$cm$ away from the hologram.
 Then, we found out the center of each circle facula on every photo, and extracted the intensity data in $x$ direction and $y$ direction through the facula center, just as Fig. 4 showing. In addition, the $ M^{2} $ values of these experimental beams were also measured. The theoretical $ M^{2} $ values of LG beams scale with their mode indices $p$ and $l$ according to $ M^{2} = 2p+l+1$.

\begin{figure}[htb]
\centerline{\includegraphics[width=7.5cm]{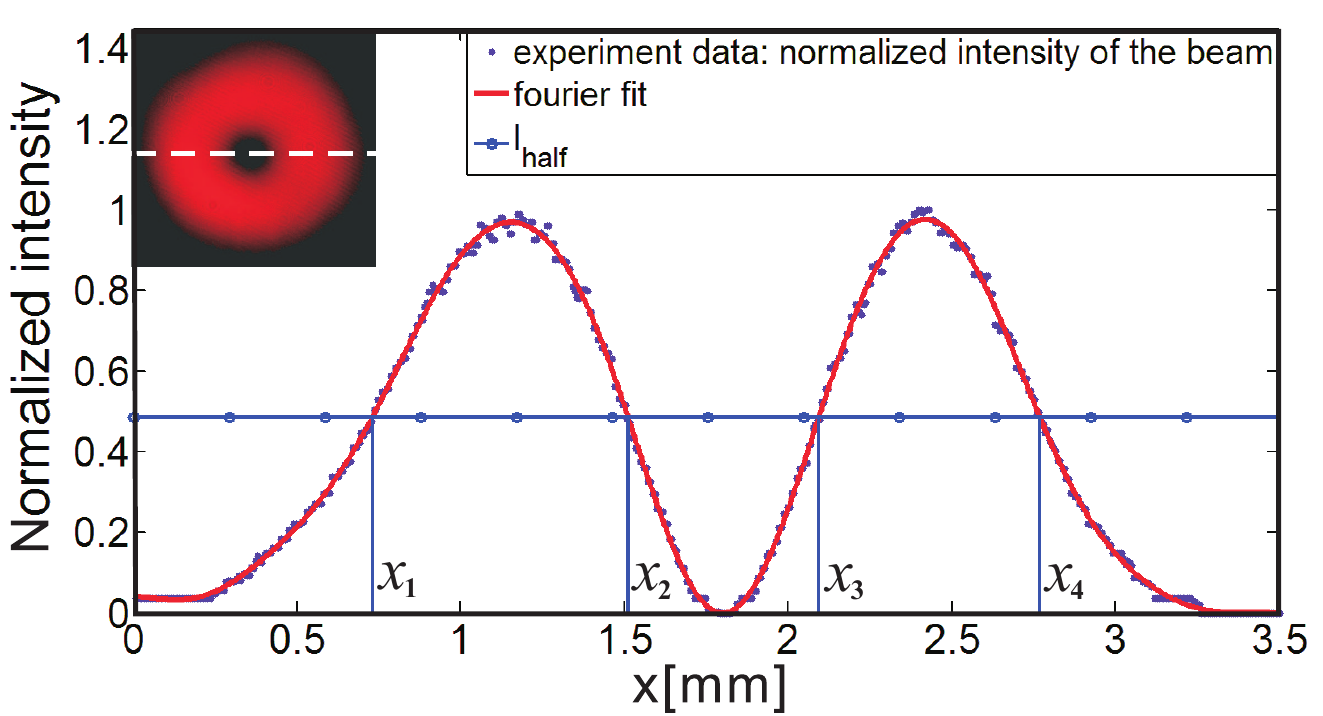}}
\caption{A far-field pattern of a beam with $p$=0 and $l$=1 and the normalized intensity distribution in the $x$ direction. We used Fourier series to fit the experimental data, and then found the \emph{$I_{half}$} line. The $x_{1}$, $x_{2}$, $x_{3}$ and $x_{4}$ are the positions that the intensity of the beam reaches \emph{$I_{half}$}. From the definition of $Q^{l}_{p}$, we could calculate the $Q^{l}_{p}$ of this beam by: $Q^{1}_{0}=\frac{x_{3}-x_{2}}{x_{4}-x_{3}+x_{2}-x_{1}}$}
\label{sy}
\end{figure}

\begin{figure*}[htb]
\centerline{\includegraphics[width=13cm]{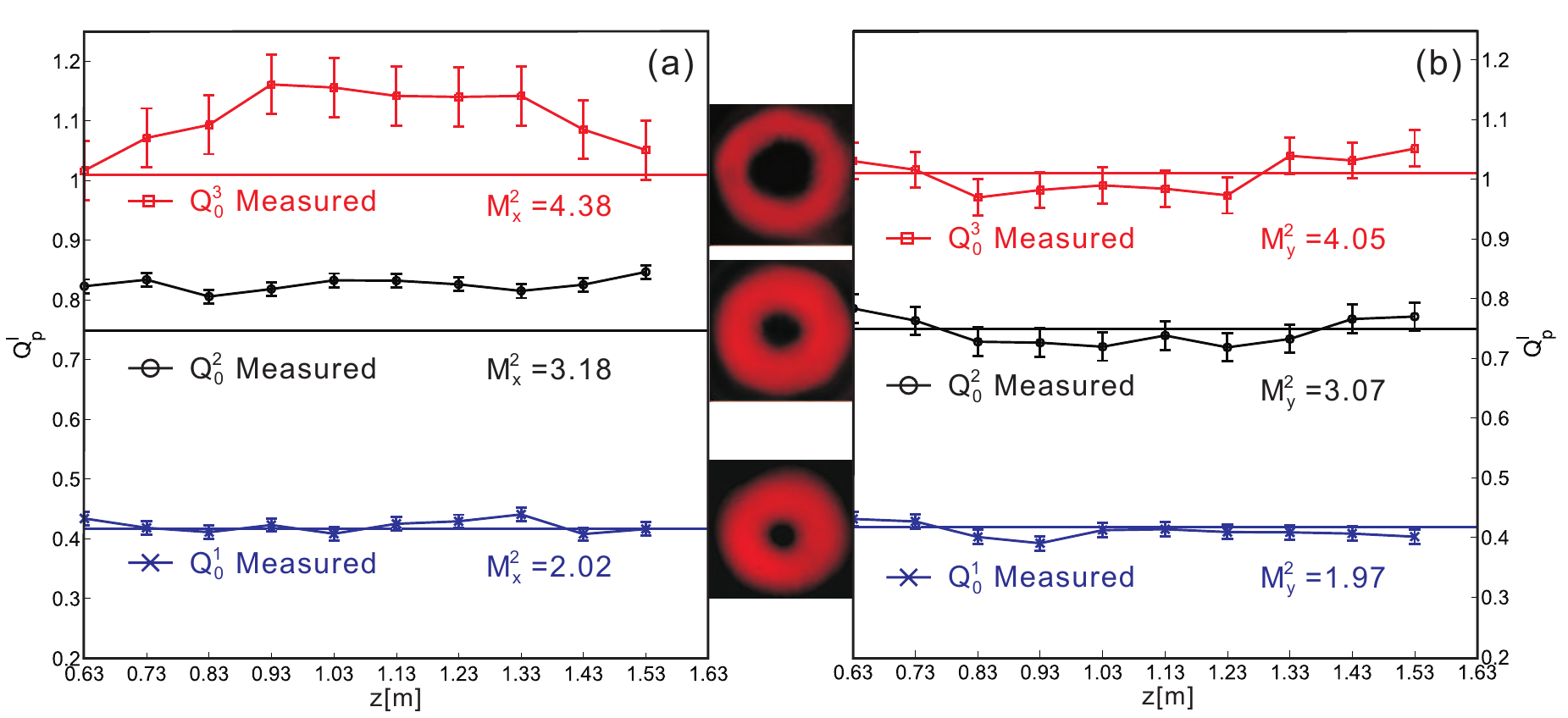}}
\caption{The measured $Q^{l}_{p}$ values in $x$ direction (a) and $y$ direction (b) of beams with $p=0$ and $l$ ranging from 1 to 3 (pictures are shown in the middle of figure (a) and (b)). The straight lines are the theoretical values. }
\label{result}
\end{figure*}

The results are shown in \reffig{result} and a comparison between theoretical values and measured $Q^{l}_{p}$ values (LG$_{01}$, LG$_{02}$ and LG$_{03}$) is made.
At least two features can be gotten from \reffig{result}. First, we can see that different LG beams possess of different $Q^{l}_{p}$ values, and $Q^{l}_{p}$ values are stable at different observation positions in a large range. It means that $Q^{l}_{p}$ can be measured conveniently at any place. To reduce the error from measurement, it is pragmatic to make an average of $Q^{l}_{p}$ values measured at different places for a beam.
Second, the $Q^{l}_{p}$ values have same trend with the $M^{2}$ values, \textit{i.e.}, when $Q^{l}_{p}$ values are close to the theoretical  values (e.g. $Q^{1}_{0}$ in $x$ and $y$ directions, $Q^{2}_{0}$ and $Q^{3}_{0}$ in $y$ direction), $M^{2}$ values are also good; when $Q^{l}_{p}$ values have a large deviation to the theoretical values (e.g. $Q^{2}_{0}$ and $Q^{3}_{0}$ in $x$ direction), $M^{2}$ values are also bad. This performance shows that $Q^{l}_{p}$ can be used to evaluate the mode quality of LG beams as $M^{2}$, while $Q^{l}_{p}$ is simpler than $M^{2}$ in practical measurement.


$Q^{l}_{p}$ also provides a convenient way to sort the OAM of LG beams. From \reffig{NC}, we can conclude that the when the mode index $p$ is certain, the value of $l$ is directly determined by the $Q^{l}_{p}$ value of the beam. Up to now, many approaches have been proposed to measure the OAM of light beams by interference \cite{interference1,interference2}, diffraction \cite{double,MPI,annular,Triangular} and atomic ensembles \cite{atom}. And some excellent works based on cascade of Mach-Zehnder interferometers \cite{M-Z} and image reformatting \cite{IT} can even measuring the OAM of single photons. What all these methods have in common is that they all identify the OAM states based on phase information. In contrast to that, the method of measuring $Q^{l}_{p}$ is based on intensity information and can be achieved more convenient by only a CCD camera.
However, due to the intensity distribution detecting, the sign of $l$ can not be determined from the $Q^{l}_{p}$ value.

In conclusion, we define a new parameter for LG model and then theoretically prove that it is a function only relating to mode indices $p$ and $l$. And the experiment results also support the theory. Furthermore, the $Q^{l}_{p}$ values of several LG beams are compared with the $M^{2}$ values in our experiment, which shows that measuring $Q^{l}_{p}$ value is a simple method  to evaluate the quality of LG mode. $Q^{l}_{p}$ can also be used to determine the OAM number of LG beams.

This work is supported by the Fundamental Research Funds for the
Central Universities, Special Prophase Project on the National Basic
Research Program of China (Grant No. 2011CB311807), and the
National Natural Science Foundation of China (Grant Nos. 11004158,
11074198, 11174233 and 11074199).


\begin{thebibliography}{99}
\bibitem{Al} L. Allen, M. W. Beijersbergen, R. J. C. Spreeuw, and J. P. Woerdman, Phys. Rev. A \textbf{45}, 8185 (1992).

\bibitem{GM}G. Molina-Terriza, J. P. Torres, and L. Torner, Phys. Rev. Lett. \textbf{88}, 013601 (2001).

\bibitem{GG} G. Gibson, J. Courtial, M. J. Padgett, M. Vasnetsov, V. Pas'ko, S. M. Barnett, and S. Franke-Arnold, Opt. Express \textbf{12}, 5448 (2004).


\bibitem{MDC} M. W. Beijersbergen, L. Allen, H. E. L. O. Vanderveen, and J. P. Woerdman, Opt. Commun. \textbf{96}, 123 (1993).


\bibitem{PP} M. W. Beijersbergen, R. P. C. Coerwinkel, M. Kristensen, and J. P. Woerdman, Opt. Commun. \textbf{112}, 321 (1994).



\bibitem{fork1} M. S. S. I. V. Basistiy, M. S. Soskin, and M. V. Vasnetsov, Opt. Commun. \textbf{119}, 604 (1995).



\bibitem{fork2} G. Brand, Am. J. Phys. \textbf{67}, 55 (1999).


\bibitem{OT}  M. E. J. Friese, J. Enger, H. Rubinsztein-Dunlop, and NR. Heckenberg, Phys. Rev. A \textbf{54}, 1593 (1996).

\bibitem{QIE}   J. T. Barreiro, T. C. Wei, and P. G. Kwiat, Nature Physics \textbf{4}, 282 (2008).

\bibitem{M2} C. Schulze, D. Flamm, M. Duparr\'{e}, and A. Forbes, Opt. Lett. \textbf{37}, 4687 (2012).

\bibitem{ISO}  Lasers and laser-related equipment - Test methods for laser beam widths, divergence angles and beam propagation ratios, BRITISH STANDARD BS EN ISO 11810-2:2007.

 \bibitem{intensity}   S. Prabhakar, A. Kumar, J. Banerji, and R. P. Singh, Opt. Lett. \textbf{36}, 4398 (2011).

 \bibitem{interference1}  M. J. Padgett, J. Arlt, N. Simpson, and L. Allen, Am. J. Phys. \textbf{64}, 77 (1996).
 
 \bibitem{interference2}  R.F. Liu, J.L. Long, F.R. Wang, Y.L. Wang, P. Zhang, H. Gao, and F.L. Li, arXiv:1305.5631.

\bibitem{double} H. I. Sztul and R. R. Alfano, Opt. Lett. \textbf{31}, 999 (2006).


\bibitem{MPI} G. C. G. Berkhout and M. W. Beijersbergen, Phys. Rev. Lett. \textbf{101}, 100801 (2008).


\bibitem{annular} C. S. Guo, L. L. Lu, and H. T. Wang, Opt. Lett. \textbf{34}, 3686 (2009).



\bibitem{Triangular} J. M. Hickmann, E. J. S. Fonseca, W. C. Soares, and S. Ch\'{a}vez-Cerda, Phys. Rev. Lett. \textbf{105}, 053904 (2010).

\bibitem{atom} L. Han, M. Cao, R. Liu, H. Liu, W. Guo, D. Wei, S. Gao, P. Zhang, H. Gao, and F. Li, Europhy. Lett. \textbf{99}, 34003 (2012).

\bibitem{M-Z} J. Leach, M. J. Padgett, S. M. Barnett, S. Franke-Arnold, and J. Courtial, Phys. Rev. Lett. \textbf{88}, 257901 (2002).

\bibitem{IT} G. C. Berkhout, M. P. Lavery, J. Courtial, M. W. Beijersbergen, and M. J. Padgett, Phys. Rev. Lett. \textbf{105}, 153601 (2010).


\end{thebibliography}
\end{document}